\documentstyle[12pt]{article}

\begin{document}

\baselineskip=14pt plus 1pt minus 1pt

\begin{center}{\large \bf 
W(5): Wobbling mode in the framework of the X(5) model
}

\bigskip\bigskip

{Dennis Bonatsos$^{\#}$\footnote{e-mail: bonat@inp.demokritos.gr},
D. Lenis$^{\#}$\footnote{e-mail: lenis@inp.demokritos.gr}, 
D. Petrellis$^{\#}$\footnote{e-mail: petrellis@inp.demokritos.gr}, 
P. A. Terziev$^\dagger$\footnote{e-mail: terziev@inrne.bas.bg} }
\bigskip

{$^{\#}$ Institute of Nuclear Physics, N.C.S.R.
``Demokritos''}

{GR-15310 Aghia Paraskevi, Attiki, Greece}

{$^\dagger$ Institute for Nuclear Research and Nuclear Energy, Bulgarian
Academy of Sciences }

{72 Tzarigrad Road, BG-1784 Sofia, Bulgaria}

\end{center}

\bigskip\bigskip
\centerline{\bf Abstract} \medskip

Using in the Bohr Hamiltonian the approximations leading to the 
Bohr and Mottelson description of wobbling motion in even nuclei, a W(5) 
model for wobbling bands, coexisting with a X(5) ground state band, is 
obtained. Separation of variables is achieved by assuming that the relevant 
potential has a sharp minimum at $\gamma_0$, which is the only parameter
entering in the spectra and B(E2) transition rates (up to overall scale 
factors). B(E2) transition rates exhibit the features expected in the 
wobbling case, while the spectrum for $\gamma=20^{\rm o}$ is in good agreement
with experimental data for $^{156}$Dy. 
 
\bigskip 

{\bf 1. Introduction} 

Nuclear wobbling motion \cite{BM} is expected to occur for triaxial nuclei at 
high angular momenta, when the angular momentum is aligned with the 
axis corresponding to the largest moment of inertia, a situation which 
classically corresponds to simple rotation without precession of the axes. 
Although wobbling motion was initially introduced for even nuclei \cite{BM}, 
it has been seen experimentally up to now (and only recently) only in odd 
nuclei ($^{163}$Lu \cite{Odegard,JensenPRL,JensenNPA}, $^{165}$Lu \cite{Schon},
$^{167}$Lu \cite{Amro}). Detailed theoretical works have been performed 
in the cranked shell model plus random phase approximation 
\cite{Shimizu,Matsu65,Matsu69}, as well as in the particle--rotor model 
\cite{Ham65,Ham67}, which naturally contain free parameters. 

In the present work we attempt a nearly parameter-free (up to overall scale 
factors) description of wobbling in even nuclei, following the methods 
developed in the E(5) \cite{IacE5}, X(5) \cite{IacX5}, Y(5) \cite{IacY5},
and Z(5) \cite{Z5} models, which correspond to the U(5)--O(6), U(5)--SU(3), 
axial--triaxial, and prolate--oblate shape phase transitions respectively.
Furthermore, the wobbling nucleus is assumed to possess a relatively rigid 
triaxial shape, as in Refs. \cite{DavFil,DavRos,Dav24}, with the potential 
having 
a sharp minimum at $\gamma=\gamma_0$. $\gamma_0$ is the only free parameter 
entering in the problem. It will be seen, however, that the results are 
changing very little with $\gamma_0$ within the region of interest.  
The path we follow is described here: 

1) We assume that the ground state band (gsb), which should be Yrast at low 
angular momentum $L$, is axial, characterized by $\gamma_0=0$. We then use for 
this purpose the X(5) gsb, which is indeed derived from the original Bohr 
Hamiltonian \cite{Bohr} after approximately separating variables for 
$\gamma=0$ \cite{IacX5}. 

2) We assume (as in Ref. \cite{Turner}) that triaxiality should appear 
at higher $L$. Starting then from the original Bohr Hamiltonian, we 
approximately separate variables following the steps of Bohr and Mottelson 
\cite{BM} in the definitive description of wobbling and keeping $\gamma$ 
close to $\gamma_0$. The resulting model, in which only $\gamma_0$ 
appears as a parameter, we call W(5). The spectrum of W(5) is measured from 
the ground state of X(5) and normalized to the first excited state of the gsb 
of X(5), in order to be directly comparable to the X(5) spectrum.        

3) The $n_w=0$ band of W(5) (where $n_w$ is the number of wobbling phonons
\cite{BM})
is found to cross the gsb of X(5) at certain $L$, depending (very weakly
within the region of interest) on $\gamma_0$. Thus the $n_w=0$ band of W(5) 
becomes Yrast beyond some specific $L$. Bands with $n_w=1$, 2, \dots exist 
at higher energies. 

4) The $n_w=0$, 1, 2 bands of W(5) are connected by intraband and interband 
B(E2) transitions which exhibit the characteristic features expected in the 
case of wobbling \cite{CastenW}. 

It is clear that the W(5) model should be tested against experiment in nuclei
of which the gsb at low $L$ appears to be close to X(5). A summary of such 
nuclei in the rare earth region is given in Ref. \cite{McCutchan}. It is 
indeed seen that existing experimental spectra on $^{156}$Dy \cite{Dy156}
correspond very well to the $n_w=0$ and $n_w=1$ bands of the W(5) model
for $\gamma_0=20^{\rm o}$. 

In Sections 2 and 3 of the present work the $\beta$-part and the $\gamma$-part
of the W(5) spectrum are derived respectively, while 
B(E2) transition rates are studied in Section 4. Numerical results are 
presented in Section 5, while Section 6 contains a brief comparison to 
experiment and in Section 7 discussion of the present results and plans for 
further work are given.  
 
{\bf 2. The $\beta$-part of the spectrum} 

The original Bohr Hamiltonian \cite{Bohr} is
\begin{equation}\label{eq:e1}
H = -{\hbar^2 \over 2B} \left[ {1\over \beta^4} {\partial \over \partial 
\beta} \beta^4 {\partial \over \partial \beta} + {1\over \beta^2 \sin 
3\gamma} {\partial \over \partial \gamma} \sin 3 \gamma {\partial \over 
\partial \gamma} - {1\over 4 \beta^2} \sum_{k=1,2,3} {Q_k^2 \over \sin^2 
\left(\gamma - {2\over 3} \pi k\right) } \right] +V(\beta,\gamma),
\end{equation}
where $\beta$ and $\gamma$ are the usual collective coordinates, while
$Q_k$ ($k=1$, 2, 3) are the components of angular momentum and $B$ is the 
mass parameter.  

Introducing \cite{IacX5} reduced energies $\epsilon=2B E/\hbar^2$ and reduced 
potentials $u=2B V/\hbar^2$, one aims at an approximate separation of 
variables by assuming that the reduced potential can be separated into 
two terms, one depending on $\beta$ and the other depending on $\gamma$, i.e.
$u(\beta,\gamma)=u(\beta)+u(\gamma)$. 

In the X(5) model \cite{IacX5}, approximate separation of variables 
is achieved by assuming that the potential $u(\gamma)$ 
has a minimum around $\gamma_0=0$, guaranteeing that $K$, the projection of 
angular momentum on the body-fixed $\hat z'$-axis, is a good quantum number. 
One then seeks solutions of the relevant Schr\"odinger equation having 
the form 
$ \Psi(\beta, \gamma, \theta_i)= \phi_K^L(\beta,\gamma) 
{\cal D}_{M,K}^L(\theta_i)$, 
where $\theta_i$ ($i=1$, 2, 3) are the Euler angles, ${\cal D}(\theta_i)$
denote Wigner functions of them, while $L$ and $M$ are the eigenvalues of 
angular momentum and the eigenvalues of the projection of angular 
momentum on the laboratory-fixed $\hat z$-axis respectively. 
In the case 
in which $u(\beta)$ is an infinite well potential
\begin{equation}\label{eq:e2} 
 u(\beta) = \left\{ \begin{array}{ll} 0 & \mbox{if $\beta \leq \beta_W$} \\
\infty  & \mbox{for $\beta > \beta_W$} \end{array} \right. ,  
\end{equation} 
the relevant differential equation is solved exactly, 
the corresponding  eigenvalues being
\begin{equation}\label{eq:e3}
\epsilon_{\beta; s,\nu} = (k_{s,\nu})^2, \qquad 
k_{s,\nu}=  {x_{s,\nu} \over \beta_W},
\end{equation}
where $x_{s,\nu}$ is the $s$-th zero of the Bessel function 
$J_\nu(k_{s,\nu}\beta)$, with \cite{IacX5,Bijker} 
\begin{equation}\label{eq:e4} 
\nu= \left( {L(L+1)-K^2 \over 3} +{9\over 4}\right)^{1/2}. 
\end{equation}
while the relevant eigenfunctions  are
\begin{equation}\label{eq:e5} 
\xi_{s,\nu}(\beta) = c_{s,\nu} \beta^{-3/2} J_\nu(k_{s,\nu} \beta), 
\end{equation}
where $c_{s,\nu}$ are  normalization constants.

In the Z(5) model \cite{Z5}, approximate separation of variables 
is achieved by assuming that the potential $u(\gamma)$ has a minimum around 
$\gamma_0=\pi/6$, guaranteeing \cite{MtVNPA} that $\alpha$, the projection of 
angular momentum on the body-fixed $\hat x'$-axis, is a good quantum number. 
One then seeks solutions of the relevant Schr\"odinger equation having 
the form 
$ \Psi(\beta, \gamma, \theta_i)= \xi_{L,\alpha}(\beta) \eta(\gamma) 
{\cal D}^L_{M,\alpha}(\theta_i)$.
In the case of $u(\beta)$ being an infinite well potential (Eq. (\ref{eq:e2}))
the relevant differential equation is solved exactly, 
the eigenvalues and eigenfunctions having the form given in Eqs. 
(\ref{eq:e3}) and (\ref{eq:e5}) respectively, with \cite{Z5}  
\begin{equation}\label{eq:e7} 
\nu = {\sqrt{4L(L+1)-3\alpha^2+9}\over 2}= 
{\sqrt{L(L+4)+3n_w(2L-n_w)+9}\over 2},   
\end{equation}
where $n_w=L-\alpha$ is the wobbling quantum number \cite{MtVNPA}. 

The X(5) and Z(5) solutions, briefly reviewed above, are obtained for 
specific values of $\gamma_0$ (0, $\pi/6$ respectively), and are valid 
for any value of the angular momentum $L$. A different approximate solution, 
which for brevity we are going to call W(5), can be obtained by following 
the steps of Bohr and Mottelson \cite{BM} for the description of 
wobbling motion. This solution will be obtained for a range of $\gamma_0$ 
values, but it will be valid only for large values of the angular momentum $L$,
which is supposed to be aligned along the axis corresponding to the largest 
moment of inertia. 

Using in Eq. (\ref{eq:e1}) the definitions 
\begin{equation}\label{eq:e11}
A_k = {1\over \sin^2\left( \gamma -{2\pi \over 3} k\right) }, \quad k=1,2,3
\end{equation}
one sees that in the region $0<\gamma < \pi/6$ one has $A_1< A_2 < A_3$. 
Therefore the largest moment of inertia corresponds to $k=1$. 
In what follows we are going to restrict ourselves to the $0< \gamma < \pi/6$
region. 

For large angular momenta $L$ aligned along the $k=1$ axis, following 
Bohr and Mottelson \cite{BM} one can see that the eigenvalues 
of the $ A_1 Q_1^2 + A_2 Q_2^2 + A_3 Q_3^2$ term in Eq. (\ref{eq:e1}) 
take the form 
\begin{equation}\label{eq:e12}
\varepsilon (n_w,L)= A_1 L(L+1) + 2 A_1 L \left(n_w+{1\over 2}\right) A_w,
\end{equation}
with
\begin{equation}\label{eq:e13}
A_w = \sqrt{ \left( {A_2\over A_1}-1\right) \left({A_3\over A_1} -1\right) }, 
\end{equation}
where $n_w$ is the number of the wobbling excitation quanta, 
for which the approximate (in the present case) relation $n_w= L-\alpha$ 
(where $\alpha$ is the projection of angular momentum on the $k=1$ 
body-fixed axis, as before) holds. Since $\alpha \approx L$ and $L$ is a 
good quantum number, $\alpha$ can be approximately treated as a good quantum 
number, too.  

It should be remembered that the approximations carried out in Ref. \cite{BM}
are valid for 
\begin{equation}\label{eq:e13a}
L >> \left( n_w+{1\over 2}\right) {1\over A_w} 
\left({A_2+A_3\over A_1}-2\right) \equiv L_0, 
\end{equation}
where $A_w$ is given by Eq. (\ref{eq:e13}). 

Using this result in the Schr\"odinger equation corresponding to 
the Hamiltonian of Eq. (\ref{eq:e1}), one can separate it into two equations 
\begin{equation} \label{eq:e14}
\left[ -{1\over \beta^4} {\partial \over \partial \beta} \beta^4 
{\partial \over \partial \beta} + {1\over 4 \beta^2} 
A_1 \left(L(L+1)+ 2 L \left(n_w+{1\over 2}\right) A_w \right)  
+u(\beta) \right] \xi_{L,n_w}(\beta) =\epsilon_\beta  \xi_{L,n_w}(\beta),
\end{equation}
\begin{equation}\label{eq:e15} 
\left[ -{1\over \langle \beta^2\rangle \sin 3\gamma} {\partial \over 
\partial \gamma}\sin 3\gamma {\partial \over \partial \gamma} 
+u(\gamma)\right] \eta(\gamma) = 
\epsilon_\gamma \eta(\gamma),
\end{equation}
where $\langle \beta^2 \rangle$ is the average of $\beta^2$ over $\xi(\beta)$, 
and $\epsilon= \epsilon_\beta +\epsilon_\gamma$. Here we assume, 
as in Refs. \cite{DavFil,DavRos,Dav24},  that the 
potential $u(\gamma)$ has a deep minimum at $\gamma=\gamma_0$, and that 
the variable $\gamma$ remains ``frozen'' at the value $\gamma_0$ 
in $A_1$ and $A_w$, appearing in Eq. (\ref{eq:e14}). Furthermore, the 
$\langle \beta^2 \rangle$ term introduces a ``hidden'' dependence on 
$s$, $L$, and $n_w$ in Eq. (\ref{eq:e15}). The approximate separation 
of the $\beta$ and $\gamma$ variables is achieved by considering an adiabatic 
approximation, as in the X(5) case \cite{IacX5,Bijker}. 

The total wave function should have the form 
\begin{equation}\label{eq:e16}
\Psi(\beta,\gamma,\theta_i) = \xi_{L,n_w}(\beta) \eta(\gamma) 
{\cal D}^L _{M,\alpha}(\theta_i), 
\end{equation}
with $\alpha=L-n_w$. 

In the case in which $u(\beta)$ is the infinite well potential of 
Eq. (\ref{eq:e2}), one can use the transformation \cite{IacX5} 
$\tilde \xi(\beta) = \beta^{3/2} \xi(\beta)$, as well as the definitions 
\cite{IacX5} $\epsilon_\beta= k_\beta^2$, $z=\beta k_\beta$, in order 
to bring Eq. (\ref{eq:e14}) into the form of a Bessel equation 
\begin{equation}\label{eq:e17}
{d^2 \tilde \xi \over d z^2} + {1\over z} {d \tilde \xi \over d z} 
+ \left[ 1 - {\nu^2 \over z^2}\right] \tilde \xi=0,
\end{equation}
with 
\begin{equation}\label{eq:e18} 
\nu = {\sqrt{A_1 L(L+1)+ 2 A_1 L \left(n_w+{1\over 2}\right) A_w +9}\over 2}.
\end{equation}
Then the boundary condition $\tilde \xi(\beta_W) =0$ 
determines the spectrum 
\begin{equation}\label{eq:e19}
\epsilon_{\beta; s,\nu} = \epsilon_{\beta; s,n_w,L} 
= (k_{s,\nu})^2, \qquad k_{s,\nu} = {x_{s,\nu}
\over \beta_W}, 
\end{equation}
and the eigenfunctions 
\begin{equation}\label{eq:e20} 
\xi_{s,\nu}(\beta) = \xi_{s,n_w,L} (\beta)= \xi_{s,\alpha,L}(\beta)= 
c_{s,\nu} \beta^{-3/2} J_\nu (k_{s,\nu} \beta), 
\end{equation}
where $x_{s,\nu}$ is the $s$th zero of the Bessel function $J_\nu(z)$, 
while the normalization constants $c_{s,\nu}$ are determined from the 
normalization condition $ \int_0^\infty \beta^4 \xi^2_{s,\nu}(\beta) 
d\beta=1$. The notation for the roots has been kept the same as in Ref. 
\cite{IacX5}, while for the energies the notation $E_{s,n_w,L}$ 
will be used. The lowest band corresponds to $s=1$, $n_w=0$ with 
$L=0$, 2, 4, \dots, while the next bands are $s=1$, $n_w=1$ with 
$L=1$, 3, 5, \dots, and $s=1$, $n_w=2$ with $L=2$, 4, 6, \dots
\cite{MtVNPA}. 

In the special case of $\gamma_0=\pi/6$ one can easily see 
from Eqs. (\ref{eq:e11}), (\ref{eq:e13}) that $A_1=1$, $A_2=A_3=4$, 
$A_w=3$. Then Eq. (\ref{eq:e18}) takes the form 
$\nu= {1\over 2} \sqrt{ L(L+4)+6 n_w L +9}$, which is in agreement 
with the corresponding Z(5) expression of Eq. (\ref{eq:e7}) up to terms 
of order $n_w^2$. 

{\bf 3. The $\gamma$-part of the spectrum} 

The $\gamma$-part of the spectrum is obtained from Eq. (\ref{eq:e15}), 
which can be simply rewritten as
\begin{equation}\label{eq:e21} 
\left[-{1\over \langle \beta^2 \rangle} \left( {\partial^2 \over \partial 
\gamma^2} + 3{\cos 3\gamma \over \sin 3\gamma} {\partial \over \partial 
\gamma}\right) +u(\gamma)\right] \eta(\gamma) = \epsilon_\gamma \eta(\gamma). 
\end{equation} 
We consider a harmonic oscillator potential having a sharp 
minimum at $\gamma =\gamma_0$ ($0<\gamma_0 \leq \pi/6$), i.e. 
\begin{equation}\label{eq:e22}
u(\gamma)= {1\over 2} c \left( \gamma-\gamma_0\right)^2 = 
{1\over 2} c \tilde \gamma^2, \qquad \tilde \gamma = \gamma -\gamma_0.
\end{equation}
The minimum is sharp as long as the constant $c$ is taken to be sufficiently 
large. 

Since we consider only small oscillations around $\gamma_0$, 
the above equation can be brought into the form 
\begin{equation}\label{eq:e23}
\left[ -{\partial^2 \over \partial \tilde \gamma^2} 
-3 \cot 3\gamma_0 {\partial \over \partial \tilde \gamma} 
+{1\over 2} c 
\langle \beta^2 \rangle \tilde \gamma^2 \right] \eta(\tilde \gamma) 
= \epsilon_{\tilde \gamma} \langle \beta^2 \rangle \eta(\tilde \gamma).
\end{equation}
Using 
\begin{equation}\label{eq:e24}
\eta(\tilde \gamma) = \tilde \eta (\tilde \gamma) 
e^{-3(\cot 3\gamma_0)\tilde\gamma/2},   
\end{equation}
Eq. (\ref{eq:e23}) is brought into the form 
\begin{equation}\label{eq:e25}
\left( -{\partial^2 \over \partial \tilde \gamma^2} +{1\over 2} c \langle 
\beta^2 \rangle \tilde \gamma^2 \right) \tilde \eta (\tilde \gamma) 
= \left( \epsilon_{\tilde \gamma} \langle \beta^2 \rangle -{9\over 4}
(\cot3\gamma_0)^2\right) \tilde \eta (\tilde \gamma),   
\end{equation} 
which is a simple harmonic oscillator equation with energy eigenvalues
\begin{equation}\label{eq:e26}
\tilde \epsilon_{\tilde \gamma} = \epsilon_{\tilde \gamma} \langle \beta^2 
\rangle -{9\over 4} (\cot3\gamma_0)^2 
= \sqrt{2 c \langle \beta^2 \rangle } \left(
n_{\tilde \gamma} +{1\over 2}\right), \qquad n_{\tilde \gamma}=0,1,2,\ldots  
\end{equation}
and eigenfunctions 
\begin{equation}\label{eq:e27}
\tilde \eta_{n_{\tilde \gamma}} ({\tilde \gamma}) 
= N_{n_{\tilde \gamma}} H_{n_{\tilde \gamma}}(b \tilde \gamma) 
e^{-b^2 \tilde \gamma^2 /2}, \qquad b=\left({ c \langle \beta^2 \rangle \over 
2}\right)^{1/4},   
\end{equation}
with normalization constant 
\begin{equation}\label{eq:e28} 
N_{n_{\tilde \gamma}} = \sqrt{ b\over \sqrt{\pi} 2^{n_{\tilde \gamma}} 
n_{\tilde \gamma}! }. 
\end{equation}
Similar potentials and solutions in the $\gamma$-variable have been 
considered in \cite{Dav24,Bohr}. 

The total energy in the case of the W(5) model is then
\begin{equation}\label{eq:e29}
E(s,n_w,L,n_{\tilde\gamma}) = E_0 + A (x_{s,\nu})^2 + B n_{\tilde \gamma}. 
\end{equation}

It should be noticed at this point that the $\gamma$-part of the spectrum 
is treated here only to lowest order approximation. Although this suffices 
for the purpose of the present work, in which only states with 
$n_{\tilde \gamma}=0$ are considered, higher approximations might be 
necessary for the detailed study of states with $n_{\tilde \gamma}\neq 0$. 
 
{\bf 4. B(E2) transition rates} 

As in the Z(5) case, the quadrupole operator is given by 
\begin{equation}\label{eq:e31}
T^{(E2)}_\mu = t \beta \left[ {\cal D}^{(2)}_{\mu,0}(\theta_i)\cos\left(\gamma 
-{2\pi\over 3}\right)+{1\over \sqrt{2}}
({\cal D}^{(2)}_{\mu,2}(\theta_i)+{\cal D}^{(2)}_{\mu,-2}(\theta_i) ) 
\sin\left(\gamma -{2\pi\over 3} \right) \right],
\end{equation}
where $t$ is a scale factor, 
while in the Wigner functions the quantum number $\alpha$ appears next 
to $\mu$, and the quantity $\gamma -2\pi/3$ in the trigonometric functions 
is obtained from $\gamma-2\pi k/3$ for $k=1$, since in the present case 
the projection $\alpha$ along the body-fixed $\hat x'$-axis is used. 
Eq. (\ref{eq:e31}) is equivalent to Eq. (3.6) of Ref. \cite{Shimizu}, 
up to a sign convention for $\gamma$. This indicates that the results 
of the present work correspond to region 2 ($-60^{\rm o} < \gamma < 0^{\rm o}$)
\cite{Shimizu} in the Lund convention. However, in the present case only the 
region $0^{\rm o} < \gamma < 30^{\rm o}$ is covered, because of the assumptions
made in Section 2.  

B(E2) transition rates are given by 
\begin{equation}\label{eq:e32}
B(E2; L_i \alpha_i \to L_f \alpha_f) ={5\over 16\pi} { |\langle L_f \alpha_f
|| T^{(E2)} || L_i \alpha_i\rangle|^2 \over (2L_i+1)}.  
\end{equation}

The symmetrized wave function reads 
\begin{equation}\label{eq:33}
\Psi(\beta,\gamma,\theta_i) = \xi_{s,n_w,L}(\beta) \eta(\tilde \gamma) 
\sqrt{ 2L+1\over 16\pi^2 (1+\delta_{\alpha,0})} ({\cal D}^{(L)}_{\mu,\alpha}
+(-1)^L {\cal D}^{(L)}_{\mu,-\alpha}) ,  
\end{equation}
where the normalization factor occurs from the standard integrals 
involving two Wigner functions \cite{Edmonds} and is the same as in Ref. 
\cite{MtVNPA}. $\alpha$ has to be an even integer \cite{MtVNPA}, 
while for $\alpha=0$ it is clear that only even values of $L$ are 
allowed, since the symmetrized wave function is vanishing otherwise. 

In the calculation of the matrix elements of Eq. (\ref{eq:e32}) the integral 
over $ \gamma$ leads to unity  [because of the normalization 
of $\eta(\tilde \gamma)$], the integral over $\beta$ takes the form 
\begin{equation}\label{eq:e34} 
I_\beta(s_i,L_i,\alpha_i,s_f,L_f,\alpha_f)= \int \beta 
\xi_{s_i,\alpha_i,L_i}(\beta) \xi_{s_f,\alpha_f,L_f}(\beta) \beta^4 d\beta,
\end{equation}
where the $\beta$ factor comes from Eq. (\ref{eq:e31}), and the $\beta^4$ 
factor comes from the volume element \cite{Bohr}, 
while the integral over the angles is calculated using the standard integrals 
involving three Wigner functions \cite{Edmonds}. The separation of the 
integrals occurs because $\eta(\tilde \gamma)$ does not depend on $\alpha$ 
or $n_w$, while in $\xi(\beta)$ only even values of $\alpha$ appear.  
The final result reads 
$$ B(E2; L_i \alpha_i \to L_f \alpha_f)  = {5\over 16\pi} t^2   
{1\over (1+\delta_{\alpha_i,0}) (1+\delta_{\alpha_f,0})}   $$
$$ \left[  \cos\left(\gamma -{2\pi \over 3} \right) 
\left\{ (L_i 2 L_f | \alpha_i 0 \alpha_f) 
+ (-1)^{L_i} (L_i 2 L_f | -\alpha_i 0 \alpha_f) \right\} \right. $$
$$ \left. + {1\over \sqrt{2}} \sin\left( \gamma -{2\pi\over 3}\right) 
\left\{ (L_i 2 L_f | \alpha_i 2 \alpha_f)+ (L_i 2 L_f | \alpha_i -2 \alpha_f)
+ (-1)^{L_i} (L_i 2 L_f | -\alpha_i 2 \alpha_f) \right\} \right]^2   $$
\begin{equation}\label{eq:e35} 
I_{\beta}^2(s_i,L_i, \alpha_i,s_f,L_f,\alpha_f) .  
\end{equation}
One can easily see that the Clebsch--Gordan coefficients (CGCs) appearing 
in this equation impose a 
$\Delta \alpha=0, \pm 2$ selection rule. Indeed, the first CGC is nonvanishing
only for $\alpha_i=\alpha_f$, while the third CGC is nonvanishing  
only if $\alpha_i+2 = \alpha_f$, and the fourth CGC is nonvanishing 
only if $\alpha_i-2=\alpha_f$. The second and fifth CGCs are nonvanishing only
if $\alpha_i+\alpha_f=0$ and $\alpha_i+\alpha_f=2$ respectively, which can be 
valid only in a few special cases. 

It is worth remarking that the first CGC in Eq. (\ref{eq:e35}) allows for 
nonvanishing quadrupole moments, in contrast to the Z(5) case \cite{Z5}, 
where quadrupole moments vanish up to this order. 

{\bf 5. Numerical results}

For low angular momentum the nucleus is expected to have $\gamma_0=0$. 
As angular momentum rises, at some point the nucleus will ``jump''
(as in Ref. \cite{Turner}) 
to the large $L$ limit corresponding to wobbling motion. As a consequence, 
the ground state band (gsb) of the nucleus should correspond to the gsb 
of X(5). The X(5) gsb should be the Yrast band up to some value of $L$, 
beyond which the $n_w=0$ wobbling band should become Yrast, while 
additional wobbling bands with $n_w=1$, 2, \dots should be seen further up
in energy.

It is therefore reasonable to measure all energies from the ground state 
of X(5) and normalize them to the lowest excitation energy of X(5). 
Therefore for the wobbling levels the ratios 
\begin{equation}\label{eq:e41} 
E_{s,n_w,L}^{W(5)} = {\epsilon_{\beta,s,n_w,L}^{W(5)} 
- \epsilon_{\beta,s=1,L=0}^{X(5)}
\over \epsilon_{\beta,s=1,L=2}^{X(5)} - \epsilon_{\beta,s=1,L=0}^{X(5)} } 
\end{equation}
will be used, while for the X(5) levels the ratios 
\begin{equation}\label{eq:e42}
E_{s,L}^{X(5)} = {\epsilon_{\beta,s,L}^{X(5)} -\epsilon_{\beta,s=1,L=0}^{X(5)}
\over \epsilon_{\beta,s=1,L=2}^{X(5)} -\epsilon_{\beta,s=1,L=0}^{X(5)} }
\end{equation} 
will be used. 

As we have seen in Section 2, the results depend on $\gamma_0$, the value 
of $\gamma$ at which the relevant potential is supposed to have a sharp 
minimum and at which the variable $\gamma$ is ``frozen'' in the 
$\beta$-equation (Eq. (\ref{eq:e14}). In what follows we are going to 
focus attention on $\gamma_0=15^{\rm o}$, which lies midway 
between the axial ($\gamma_0=0$) and maximally triaxial 
($\gamma_0=30^{\rm o}$) cases, and on $\gamma_0=20^{\rm o}$, which has 
been found of interest \cite{JensenNPA} in the framework of 
``Ultimate Cranker'' \cite{Bengtsson}  calculations.  

Spectra for the lowest wobbling bands for $\gamma = 15^{\rm o}$ and 
$\gamma=20^{\rm o}$ are shown in Table 1, where the gsb 
of X(5) is also shown for comparison. In Fig. 1 several levels of the 
$n_w=0$ and $n_w=1$ bands are plotted as a function of $\gamma_0$. It is seen 
that the $\gamma_0$-dependence of the energy levels is rather flat within the 
region of interest ($10^{\rm o} < \gamma_0 < 30^{\rm o}$),  therefore the two 
$\gamma_0$ values shown in Table 1 suffice. 

In Table 1 the restrictions imposed by the condition of Eq. (\ref{eq:e13a}) 
are reminded by reporting the appropriate $L_0$ values and by putting in 
parentheses the energies of the levels below this limit. 

In Table 1 we remark that the $n_w=0$ band with $\gamma_0=15^{\rm o}$ crosses 
the X(5) gsb above $L=12$ and becomes Yrast from $L=14$ up, while 
the $n_w=0$ band with $\gamma_0=20^{\rm o}$ crosses the X(5) gsb above $L=8$ 
and becomes Yrast from $L=10$ up. The angular momentum $L$ at which the 
bandcrossing of the X(5) gsb and the $n_w=0$ band occurs does not 
depend on any free parameter, but only on $\gamma_0$, the relevant 
dependence being shown in Fig. 2~. We remark that this $L$ changes very 
little in the region $15^{\rm o} < \gamma_0 < 30^{\rm o}$. 

Intraband B(E2) transition rates of the wobbling bands, as well as interband 
B(E2) transition rates among wobbling bands are shown in Table 2, again 
for $\gamma_0=15^{\rm o}$ and $\gamma_0=20^{\rm o}$,  normalized 
to the B(E2) transition rate between the two lowest $n_w=0$ states.  
Some clear features are observed: 

1) Intraband $L+2\to L$ transitions are strong, while intraband $L\to L$ 
transitions are weak. For interband $(n_w=2)\to (n_w=0)$ transitions 
the opposite situation appears, i.e. they are strong for $L\to L$ and weak 
for $L+2 \to L$. 

2) Interband $(n_w=1)\to (n_w=0)$ transitions are strong for $L \to L+1$
and weak for $L\to L-1$, while for interband $(n_w=1)\to (n_w=2)$ 
transitions the opposite picture appears, i.e. they are strong for 
$L\to L-1$ and weak for $L\to L+1$. 

It is worth comparing these results to the main features expected to be 
exhibited by B(E2)s in wobbling bands \cite{CastenW}.

1) In region 2 of the Lund convention, which corresponds to the present case, 
the interband $(n_w=1) \to (n_w=0)$ transitions are expected to be strong 
for $L\to L+1$ and weak for $L\to L-1$ \cite{Shimizu}. This is exactly 
the situation seen in Table 2. 

2) The ratio 
\begin{equation}\label{eq:e43}
{B(E2)_{out} \over B(E2)_{in} } = { B[E2; L_1 \to (L+1)_0] \over 
B[E2; L_1 \to (L-2)_1]} ,
\end{equation}
where the notation $L_{n_w}$ is used, is expected \cite{CastenW} to be 
of the order 0.2--0.3, i.e. much larger than what is expected for typical 
interband transitions. The $(n_w=1)\to (n_w=0)$ transitions in Table 2
do exhibit this behaviour. 

3) The $B(E2)_{out} = B[E2; L_1 \to (L+1)_0]$ values are expected 
to go as $1/L$ and not as $1/L^2$ \cite{CastenW}. The results in Table 2 
do exhibit this feature.  

{\bf 6. Comparison to experiment}

The results of the present approach should be compared to experimental data 
for nuclei which exhibit a X(5) ground state band. Rare earth nuclei 
with this property have been summarized in Ref. \cite{McCutchan}.
In Table 3 we see that the gsb of $^{156}$Dy \cite{Dy156} is in good agreement
with the X(5) predictions, while two additional bands (the K and M bands of 
Ref. \cite{Dy156}) respectively) correspond very well 
to the $n_w=0$ and $n_w=1$ W(5) bands with $\gamma_0=20^{\rm o}$, although
no rms fitting with respect to $\gamma_0$ has been performed. 
It is therefore of great interest 
to measure intraband and interband B(E2) transitions for these bands, in order
to see if they will exhibit the characteristic wobbling features mentioned in 
Section 5.  It should be noticed that the spin assignments in the M band 
are based on the only transition seen experimentally \cite{Dy156,Kondev}
to connect a level 
of this band to the $L=30$ level of the K band. Since in the $(n_w=1) \to 
(n_w=0)$ case the transitions $L\to L+1$ are the strong ones, as found 
in Section 5, it is assumed that the level of the M band from which this 
transition starts is the $L=29$ level.  

It should be remembered at this point that in general $\gamma$-soft models 
involving $\gamma$-flu\-ctu\-a\-tions and $\gamma$-rigid models corresponding 
to large rigid triaxiality lead to similar results for many observables
\cite{CastenW,Casten}, when $\gamma_{rms}$ of the former equals 
$\gamma_{rigid}$ of the latter. 
Therefore the agreement of the present model to experiment,
if proved for B(E2)s, too, offers indeed evidence for triaxiality but not 
necessarily for $\gamma$-rigid behaviour, although $\gamma$-rigidity was among
the assumptions which led to this model. 

{\bf 7. Discussion}

In summary, a W(5) model describing the wobbling bands coexisting with 
a X(5) ground state band in even nuclei has been introduced. 
Separation of variables is achieved by assuming that the potential has a 
sharp minimum at $\gamma=\gamma_0$. The model 
predictions for given value of $\gamma_0$ are parameter-free (up to overall 
scale factors). The W(5) predictions for wobbling bands for 
$\gamma_0 =20^{\rm o}$ are in good agreement with experimental spectra for 
$^{156}$Dy, the ground state band of which is described satisfactorily 
by X(5), while the W(5) predictions for intraband and interband B(E2) 
transition probabilities exhibit the features expected for wobbling bands.  
A characteristic feature of the model is that the $n_w=0$ wobbling band 
is not coinciding with the gsb, but with the superband crossing the gsb. 

Concerning further work, the following comments can be made: 

1) There exist nuclei ($^{160}$Yb, $^{158}$Er), the ground state bands of 
which are descibed well by the X(5)-$\beta^4$ and X(5)-$\beta^6$ models 
\cite{X5} respectively. These models correspond to the use of $\beta^4$ and 
$\beta^6$ potentials in the X(5) framework, leading to $R_4=E(4)/E(2)$ ratios 
of 2.769 and 2.824 respectively. It is worth using the $\beta^4$ and 
$\beta^6$ potentials in the W(5) framework as well, in order to examine 
if the parameter-free (up to $\gamma_0$) predictions for wobbling bands which 
will occur in these models agree with experiment. 

2) The $\beta$-equation [Eq. (\ref{eq:e14})] obtained above in the W(5) 
framework is also exactly soluble \cite{Elliott,Rowe} for the Davidson 
potentials \cite{Dav} 
\begin{equation}\label{eq:e51}
u(\beta)= \beta^2 + {\beta_0^4 \over \beta^2}, 
\end{equation}
where $\beta_0$ is the position of the minimum of the potential. 
In analogy to earlier work in the E(5) and X(5) frameworks \cite{varPLB}
it is expected that $\beta_0=0$ will correspond to a ``wobbling vibrator'', 
while $\beta_0\to\infty$ will lead to the original wobbling rotator of
Ref. \cite{BM}. 

3) Using the variational procedure developed recently in the E(5) and X(5) 
frameworks \cite{varPLB}, one should be able to prove that the W(5) model
can be obtained from the Davidson potentials by maximizing the rate 
of change of various measures of collectivity with respect to the 
parameter $\beta_0$, thus proving that W(5) corresponds to the critical 
point symmetry of the transition from a ``wobbling vibrator'' to 
a wobbling rotator. 

Work in these directions is in progress.  

{\bf Acknowledgements} 

The authors are thankful to Jean Libert (Orsay) and 
Werner Scheid (Giessen) for illuminating discussions.

\bigskip 

\centerline{\bf Figure captions}

\bigskip
{\bf Fig. 1} Energy levels with various $L$ (and $s=1$) of the W(5) model 
plotted versus $\gamma_0$, the position at which the potential has a 
sharp minimum. a) The $n_w=0$ band. b) The $n_w=1$ band.  See Section 5 
for further discussion. 

\medskip 
{\bf Fig. 2} Angular momentum $L$ up to which the X(5) ground state band is 
Yrast, while beyond it the $n_w=0$ W(5) band becomes Yrast, as a function 
of $\gamma_0$, the position at which the potential has a sharp minimum. 
See Section 5 for further discussion.   

\newpage 
%%%%%%%%%%%%%%%%%%%%%%%%%%%%%%%%%%%%%%%%%%%%%%%%%%%%%%%%%%%%%%%%%%%%%%
%%%%%%%%%%%%%%%%%%% Table 1 %%%%%%%%%%%%%%%%%%%%%%%%%%%%%%%%%%%%%%%%

\begin{table}

\caption{Energy levels of the W(5) model (with $s=1$), measured from the 
ground state 
of X(5) and normalized to the first excited state of X(5). The ground state 
band of X(5), normalized in the same way, is shown for comparison. 
See Sections 2 and 5 for further details.   
}

\bigskip

\begin{tabular}{ r r r r r r | r r r r r}
\hline
      & X(5) & W(5) & W(5) & W(5) & W(5) &  & W(5) & W(5) & W(5) & W(5) \\
$n_w$ &     &  0  &   0   & 2     &  2    &    & 1   &  1  &  3  &   3     \\
$\gamma_0$& & $15^{\rm o}$ & $20^{\rm o}$ & $15^{\rm o}$ & $20^{\rm o}$ &  &
              $15^{\rm o}$ & $20^{\rm o}$ & $15^{\rm o}$ & $20^{\rm o}$ \\
$L_0$ & & 2.061 & 1.378 & 10.307 & 6.891 & & 6.184 & 4.134 & 14.429 & 
9.647 \\
$L$&        &        &        &        &        &$L$ &        &        &
        &        \\
\hline
   &        &        &        &        &        &    &        &        &
        &        \\
 0 &  0.000 &(0.000) &(0.000) &        &        &    &        &        &
        &        \\
 2 &  1.000 &(1.614) &  1.513 & (4.293)& (3.967)&  3 & (4.535)& (4.222)&
        &        \\
 4 &  2.904 &  3.729 &  3.531 & (8.276)& (7.692)&  5 & (7.744)&  7.253 & 
(12.766)&(11.850)\\
 6 &  5.430 &  6.265 &  5.964 &(12.427)&(11.593)&  7 & 11.257 & 10.586 &
(17.790)&(16.555)\\
 8 &  8.483 &  9.195 &  8.780 &(16.841)& 15.755 &  9 & 15.099 & 14.241 &
(23.062)&(21.506)\\
10 & 12.027 & 12.504 & 11.964 &(21.555)& 20.210 & 11 & 19.281 & 18.226 &
(28.618)& 26.734 \\
12 & 16.041 & 16.184 & 15.507 & 26.585 & 24.973 & 13 & 23.806 & 22.543 &
(34.475)& 32.256 \\
14 & 20.514 & 20.227 & 19.401 & 31.940 & 30.052 & 15 & 28.675 & 27.194 &
 40.645 & 38.082 \\
16 & 25.437 & 24.630 & 23.642 & 37.624 & 35.450 & 17 & 33.887 & 32.177 &
 47.134 & 44.218 \\
18 & 30.804 & 29.389 & 28.226 & 43.641 & 41.170 & 19 & 39.444 & 37.492 &
 53.945 & 50.667 \\
20 & 36.611 & 34.501 & 33.151 & 49.991 & 47.212 & 21 & 45.343 & 43.138 &
 61.083 & 57.431 \\
22 & 42.853 & 39.962 & 38.413 & 56.676 & 53.577 & 23 & 51.584 & 49.113 &
 68.548 & 64.511 \\
24 & 49.528 & 45.772 & 44.011 & 63.695 & 60.266 & 25 & 58.166 & 55.419 &
 76.342 & 71.910 \\
26 & 56.633 & 51.927 & 49.942 & 71.049 & 67.278 & 27 & 65.087 & 62.052 &
 84.464 & 79.626 \\
28 & 64.166 & 58.427 & 56.205 & 78.737 & 74.613 & 29 & 72.348 & 69.011 &
 92.917 & 87.661 \\
30 & 72.124 & 65.270 & 62.798 & 86.759 & 82.270 & 31 & 79.947 & 76.297 &
101.699 & 96.014 \\
32 & 80.505 & 72.454 & 69.720 & 95.116 & 90.250 & 33 & 87.883 & 83.908 &
110.811 &104.686 \\
34 & 89.309 & 79.978 & 76.970 &103.805 & 98.551 & 35 & 96.156 & 91.843 &
120.253 &113.675 \\
36 & 98.533 & 87.841 & 84.546 &112.828 &107.173 & 37 &104.765 &100.102 &
130.024 &122.983 \\
38 &108.176 & 96.042 & 92.448 &122.183 &116.115 & 39 &113.708 &108.684 &
140.126 &132.609 \\
40 &118.237 &104.580 &100.675 &131.870 &125.378 & 41 &122.986 &117.588 &
150.556 &142.552 \\
42 &128.715 &113.454 &109.225 &141.888 &134.960 & 43 &132.597 &126.813 &
161.315 &152.813 \\ 
44 &139.608 &122.664 &118.098 &152.238 &144.862 & 45 &142.541 &136.360 &
172.404 &163.391 \\
46 &150.917 &132.208 &127.294 &162.919 &155.082 & 47 &152.818 &146.226 &
183.821 &174.285 \\
48 &162.639 &142.086 &136.810 &173.930 &165.621 & 49 &163.427 &156.412 &
195.566 &185.496 \\
50 &174.774 &152.297 &146.648 &185.270 &176.477 & 51 &174.367 &166.917 &
207.639 &197.023 \\
52 &187.321 &162.840 &156.806 &196.941 &187.651 & 53 &185.637 &177.741 &
220.040 &208.867 \\
54 &200.280 &173.715 &167.284 &208.940 &199.142 & 55 &197.238 &188.884 &
232.768 &221.025 \\
56 &213.650 &184.922 &178.080 &221.268 &210.950 & 57 &209.169 &200.344 &
245.823 &233.499 \\
58 &227.430 &196.459 &189.195 &233.925 &223.074 & 59 &221.430 &212.121 &
259.206 &246.288 \\
60 &241.620 &208.327 &200.629 &246.909 &235.514 & 61 &234.019 &224.215 &
272.915 &259.391 \\
   &        &        &        &        &        &    &        &        &
        &        \\
\hline
\end{tabular}
\end{table}

\newpage 
%%%%%%%%%%%%%%%%%%%%%%%%%%%%%%%%%%%%%%%%%%%%%%%%%%%%%%%%%%%%%%%%%%%%%%
%%%%%%%%%%%%%%%%%%% Table 2 %%%%%%%%%%%%%%%%%%%%%%%%%%%%%%%%%%%%%%%%

\begin{table}

\caption{B(E2) transition rates of the W(5) model (connecting states 
with $s=1$), normalized 
to the transition between the two lowest states, B(E2;$2_0\to 0_0$). 
See Sections 4 and 5 for further details. 
}

\bigskip

\begin{tabular}{ r r r r | r r r r | r r r r}
\hline
  &$\gamma_0$& $15^{\rm o}$ & $20^{\rm o}$ &  &  & $15^{\rm o}$ & 
$20^{\rm o}$ &  &  & $15^{\rm o}$ & $20^{\rm o}$ \\
 $L^{(i)}_{n_w}$ & $L^{(f)}_{n_w}$ &      &     & 
 $L^{(i)}_{n_w}$ & $L^{(f)}_{n_w}$ &      &     &
 $L^{(i)}_{n_w}$ & $L^{(f)}_{n_w}$ &      &     \\
\hline
       &       &       &       &       &       &       &       &        &
       &       &       \\
 $2_0$ & $0_0$ & 1.000 & 1.000 &       &       &       &       &        &
       &       &       \\
 $4_0$ & $2_0$ & 1.606 & 1.597 & $4_2$ & $2_2$ & 0.785 & 0.777 &
 $5_1$  & $3_1$ & 1.268 & 1.256 \\   
 $6_0$ & $4_0$ & 2.228 & 2.213 & $6_2$ & $4_2$ & 1.072 & 1.061 &
 $7_1$  & $5_1$ & 1.889 & 1.872 \\
 $8_0$ & $6_0$ & 2.665 & 2.647 & $8_2$ & $6_2$ & 1.635 & 1.620 &
 $9_1$  & $7_1$ & 2.349 & 2.330 \\
$10_0$ & $8_0$ & 2.998 & 2.980 &$10_2$ & $8_2$ & 2.081 & 2.063 &
 $11_1$ & $9_1$ & 2.705 & 2.685 \\
$12_0$ &$10_0$ & 3.265 & 3.247 &$12_2$ &$10_2$ & 2.440 & 2.421 &
 $13_1$ &$11_1$ & 2.992 & 2.971 \\ 
$14_0$ &$12_0$ & 3.485 & 3.468 &$14_2$ &$12_2$ & 2.734 & 2.715 &
 $15_1$ &$13_1$ & 3.228 & 3.208 \\
$16_0$ &$14_0$ & 3.672 & 3.655 &$16_2$ &$14_2$ & 2.981 & 2.962 &
 $17_1$ &$15_1$ & 3.429 & 3.409 \\
$18_0$ &$16_0$ & 3.832 & 3.816 &$18_2$ &$16_2$ & 3.192 & 3.173 &
 $19_1$ &$17_1$ & 3.601 & 3.582 \\
$20_0$ &$18_0$ & 3.972 & 3.956 &$20_2$ &$18_2$ & 3.374 & 3.356 &
 $21_1$ &$19_1$ & 3.752 & 3.734 \\
       &       &       &       &       &       &       &       &
        &       &       &       \\
 $2_0$ & $2_0$ & 0.114 & 0.049 & $2_2$ & $2_2$ & 0.131 & 0.056 &
        &       &       &       \\ 
 $4_0$ & $4_0$ & 0.228 & 0.098 & $4_2$ & $4_2$ & 0.021 & 0.009 &
 $5_1$  & $5_1$ & 0.047 & 0.020 \\
 $6_0$ & $6_0$ & 0.307 & 0.132 & $6_2$ & $6_2$ & 0.003 & 0.001 &
 $7_1$  & $7_1$ & 0.118 & 0.051 \\
 $8_0$ & $8_0$ & 0.366 & 0.158 & $8_2$ & $8_2$ & 0.036 & 0.016 &
 $9_1$  & $9_1$ & 0.184 & 0.079 \\
$10_0$ &$10_0$ & 0.413 & 0.178 &$10_2$ &$10_2$ & 0.084 & 0.036 &
 $11_1$ &$11_1$ & 0.240 & 0.103 \\
$12_0$ &$12_0$ & 0.451 & 0.194 &$12_2$ &$12_2$ & 0.132 & 0.057 &
 $13_1$ &$13_1$ & 0.288 & 0.124 \\
$14_0$ &$14_0$ & 0.482 & 0.208 &$14_2$ &$14_2$ & 0.178 & 0.076 &
 $15_1$ &$15_1$ & 0.329 & 0.142 \\
$16_0$ &$16_0$ & 0.510 & 0.220 &$16_2$ &$16_2$ & 0.219 & 0.094 &
 $17_1$ &$17_1$ & 0.365 & 0.157 \\
$18_0$ &$18_0$ & 0.533 & 0.230 &$18_2$ &$18_2$ & 0.257 & 0.111 &
 $19_1$ &$19_1$ & 0.397 & 0.171 \\
$20_0$ &$20_0$ & 0.554 & 0.239 &$20_2$ &$20_2$ & 0.291 & 0.125 &
 $21_1$ &$21_1$ & 0.424 & 0.183 \\
       &       &       &       &       &       &       &       &
        &       &       &       \\
\hline
       &       &       &       &       &       &       &       &
       &       &       &       \\
 $2_2$ & $2_0$ & 1.657 & 1.647 & $3_1$ & $4_0$ & 1.267 & 1.256 &
 $3_1$ & $2_2$ & 2.292 & 2.267 \\    
 $4_2$ & $4_0$ & 0.352 & 0.350 & $5_1$ & $6_0$ & 0.988 & 0.980 &
 $5_1$ & $4_2$ & 1.355 & 1.341 \\
 $6_2$ & $6_0$ & 0.199 & 0.198 & $7_1$ & $8_0$ & 0.819 & 0.814 &
 $7_1$ & $6_2$ & 1.290 & 1.279 \\ 
 $8_2$ & $8_0$ & 0.130 & 0.129 & $9_1$ &$10_0$ & 0.705 & 0.700 &
 $9_1$ & $8_2$ & 1.187 & 1.177 \\
$10_2$ &$10_0$ & 0.092 & 0.092 &$11_1$ &$12_0$ & 0.621 & 0.617 &
$11_1$ &$10_2$ & 1.086 & 1.078 \\
$12_2$ &$12_0$ & 0.069 & 0.069 &$13_1$ &$14_0$ & 0.556 & 0.554 &
$13_1$ &$12_2$ & 0.998 & 0.991 \\
$14_2$ &$14_0$ & 0.054 & 0.054 &$15_1$ &$16_0$ & 0.505 & 0.503 &
$15_1$ &$14_2$ & 0.921 & 0.916 \\
$16_2$ &$16_0$ & 0.043 & 0.043 &$17_1$ &$18_0$ & 0.463 & 0.461 &
$17_1$ &$16_2$ & 0.855 & 0.851 \\
$18_2$ &$18_0$ & 0.036 & 0.036 &$19_1$ &$20_0$ & 0.428 & 0.426 &
$19_1$ &$18_2$ & 0.798 & 0.794 \\
$20_2$ &$20_0$ & 0.030 & 0.030 &$21_1$ &$22_0$ & 0.398 & 0.397 &
$21_1$ &$20_2$ & 0.748 & 0.744 \\
       &       &       &       &       &       &       &       &
        &       &      &       \\  
 $2_2$ & $0_0$ & 0.072 & 0.031 & $3_1$ & $2_0$ & 0.149 & 0.064 &
 $3_1$  & $4_2$ & 0.164 & 0.070 \\   
 $4_2$ & $2_0$ & 0.049 & 0.021 & $5_1$ & $4_0$ & 0.152 & 0.065 &
 $5_1$  & $6_2$ & 0.216 & 0.093 \\ 
 $6_2$ & $4_0$ & 0.029 & 0.012 & $7_1$ & $6_0$ & 0.140 & 0.060 &
 $7_1$  & $8_2$ & 0.220 & 0.095 \\
 $8_2$ & $6_0$ & 0.019 & 0.008 & $9_1$ & $8_0$ & 0.127 & 0.055 &
 $9_1$  &$10_2$ & 0.212 & 0.091 \\
$10_2$ & $8_0$ & 0.013 & 0.006 &$11_1$ &$10_0$ & 0.116 & 0.050 &
$11_1$  &$12_2$ & 0.200 & 0.086 \\
$12_2$ &$10_0$ & 0.010 & 0.004 &$13_1$ &$12_0$ & 0.107 & 0.046 &
$13_1$  &$14_2$ & 0.188 & 0.081 \\
$14_2$ &$12_0$ & 0.008 & 0.003 &$15_1$ &$14_0$ & 0.098 & 0.042 &
$15_1$  &$16_2$ & 0.176 & 0.076 \\
$16_2$ &$14_0$ & 0.006 & 0.003 &$17_1$ &$16_0$ & 0.091 & 0.039 &
$17_1$  &$18_2$ & 0.166 & 0.071 \\
$18_2$ &$16_0$ & 0.005 & 0.002 &$19_1$ &$18_0$ & 0.085 & 0.037 &
$19_1$  &$20_2$ & 0.156 & 0.067 \\
$20_2$ &$18_0$ & 0.004 & 0.002 &$21_1$ &$20_0$ & 0.080 & 0.035 &
$21_1$  &$22_2$ & 0.148 & 0.064 \\
\hline
\end{tabular}
\end{table}

\newpage 
%%%%%%%%%%%%%%%%%%%%%%%%%%%%%%%%%%%%%%%%%%%%%%%%%%%%%%%%%%%%%%%%%%%%%%
%%%%%%%%%%%%%%%%%%% Table 3 %%%%%%%%%%%%%%%%%%%%%%%%%%%%%%%%%%%%%%%%

\begin{table}

\caption{Comparison of the X(5) predictions for the ground state band and the 
W(5) predictions (with $\gamma_0= 20^{\rm o}$)
for the $n_w=0$, 1 bands (with $s=1$) to experimental data 
for $^{156}$Dy \cite{Dy156}. See Section 6 for further discussion. 
}

\bigskip

\begin{tabular}{ r r r | r r r | r r r | r r r }
\hline
       & X(5) & exp.&  & W(5) & exp. &   & W(5) & exp. &  & W(5) & exp. \\
$ n_w$ &     &     &      & 0 &     &  & 0 &    &   & 1 &   \\
$L$  &     &     & $L$  &   &   & $L$  &   &   & $L$  &  &     \\
\hline
   &   &     &   &     &   &     &    &  & & &      \\
 0&  0.000 &  0.000 & 14& 19.401 & 22.254 & 38& 92.448 & 92.686 & 
29 & 69.011 & 70.349 \\
 2&  1.000 &  1.000 & 16& 23.642 & 25.396 & 40&100.675 &100.785 & 
31 & 76.297 & 77.150 \\
 4&  2.904 &  2.934 & 18& 28.226 & 29.221 & 42&109.225 &108.839 & 
33 & 83.908 & 84.300 \\
 6&  5.430 &  5.592 & 20& 33.151 & 33.647 & 44&118.098 &117.378 & 
35 & 91.843 & 91.660 \\ 
 8&  8.483 &  8.823 & 22& 38.413 & 38.617 & 46&127.294 &125.920 & 
37 &100.102 & 99.339 \\
10& 12.027 & 12.521 & 24& 44.011 & 44.060 & 48&136.810 &135.116 & 
39 &108.684 &107.425 \\
12& 16.041 & 16.592 & 26& 49.942 & 49.923 & 50&146.648 &144.828 & 
41 &117.588 &115.954 \\ 
14& 20.514 & 20.961 & 28& 56.205 & 56.172 & 52&156.806 &155.491 & 
43 &126.813 &125.107 \\
16& 25.437 & 25.574 & 30& 62.798 & 62.792 & 54&167.284 &166.930 & 
45 &136.360 &135.008 \\
18& 30.804 & 30.327 & 32& 69.720 & 69.763 & 56&178.080 &179.400 & 
47 &146.226 & 145.184 \\
20& 36.611 & 35.269 & 34& 76.970 & 77.070 & 58&189.195 &193.366 & 
49 &156.412 & 156.144 \\
22& 42.853 & 40.451 & 36& 84.546 & 84.711 &   &        &        & 
51 &166.917 & 168.716 \\
   &   &     &   &     &    &   &     &   &   &   &     \\
\hline
\end{tabular}
\end{table}


\begin{thebibliography}{99}

\bibitem{BM}
A. Bohr and B. R. Mottelson, Nuclear Structure, Benjamin, New York, 
1975, Vol. II, section 4-5e.  

\bibitem{Odegard}
S. W. \O deg\aa rd et al., Phys. Rev. Lett. 86 (2001) 5866. 

\bibitem{JensenPRL}
D. R. Jensen et al., Phys. Rev. Lett. 89 (2002) 142503. 

\bibitem{JensenNPA}
D. R. Jensen et al., Nucl. Phys. A 703 (2002) 3. 

\bibitem{Schon}
G. Sch\"onwa\ss er et al., Phys. Lett. B 552 (2003) 9. 

\bibitem{Amro}
H. Amro et al., Phys. Lett. B 553 (2003) 197. 

\bibitem{Shimizu}
Y. R. Shimizu and M. Matsuzaki, Nucl. Phys. A 588 (1995) 559. 

\bibitem{Matsu65}
M. Matsuzaki, Y. R. Shimizu, and K. Matsuyanagi, Phys. Rev. C 65 (2002) 
041303. 

\bibitem{Matsu69}
M. Matsuzaki, Y. R. Shimizu, and K. Matsuyanagi, Phys. Rev. C 69 (2004)
034325. 

\bibitem{Ham65}
I. Hamamoto, Phys. Rev. C 65 (2002) 044305. 

\bibitem{Ham67}
I. Hamamoto, Phys. Rev. C 67 (2003) 014319. 

\bibitem{IacE5} 
F. Iachello, Phys. Rev. Lett. 85 (2000) 3580. 

\bibitem{IacX5}
F. Iachello, Phys. Rev. Lett. 87 (2001) 052502. 

\bibitem{IacY5}
F. Iachello, Phys. Rev. Lett. 91 (2003) 132502. 

\bibitem{Z5}
D. Bonatsos, D. Lenis, D. Petrellis, and P. A. Terziev, Phys. Lett. B, 
in press. nucl-th/0402087. 

\bibitem{DavFil}
A. S. Davydov and G. F. Filippov, Nucl. Phys. 8 (1958) 237. 

\bibitem{DavRos} 
A. S. Davydov and V. S. Rostovsky, Nucl. Phys. 12 (1959) 58. 

\bibitem{Dav24}
A. S. Davydov, Nucl. Phys. 24 (1961) 682. 

\bibitem{Bohr}
A. Bohr, Mat. Fys. Medd. K. Dan. Vidensk. Selsk.  26 (1952) no. 14. 

\bibitem{Turner}
R. J. Turner and T. Kishimoto, Nucl. Phys. A 217 (1971) 317. 

\bibitem{CastenW}
R. F. Casten, E. A. McCutchan, N. V. Zamfir, C. W. Beausang, and J.-Y. Zhang, 
Phys. Rev. C 67 (2003) 064306. 

\bibitem{McCutchan}
E. A. McCutchan et al., Phys. Rev. C 69 (2004) 024308. 

\bibitem{Dy156}
C. W. Reich, Nucl. Data Sheets 99 (2003) 753.  

\bibitem{Bijker}
R. Bijker, R. F. Casten, N. V. Zamfir, and E. A. McCutchan, Phys. Rev. C 68
(2003) 064304. 

\bibitem{MtVNPA}
J. Meyer-ter-Vehn, Nucl. Phys. A 249 (1975) 111. 

\bibitem{Edmonds}
A. R. Edmonds, Angular Momentum in Quantum Mechanics, Princeton
University Press, Princeton, 1957. 

\bibitem{Bengtsson}
T. Bengtsson, Nucl. Phys. A 496 (1989) 56; 512 (1990) 124. 

\bibitem{Kondev}
F. G. Kondev et al., Phys. Lett. B 437 (1998) 35. 

\bibitem{Casten}
R. F. Casten, Nuclear Structure from a Simple Perspective, Oxford University 
Press, Oxford, 1990. 

\bibitem{X5}
D. Bonatsos, D. Lenis, N. Minkov, P. P. Raychev, and P. A. Terziev, 
Phys. Rev. C 69 (2004) 014302. 

\bibitem{Elliott} 
J. P. Elliott, J. A. Evans, and P. Park, Phys. Lett. B 169 (1986) 309.

\bibitem{Rowe}
D. J. Rowe and C. Bahri, J. Phys. A 31 (1998) 4947. 

\bibitem{Dav}
P. M. Davidson, Proc. R. Soc. 135 (1932) 459.  

\bibitem{varPLB}
D. Bonatsos, D. Lenis, N. Minkov, D. Petrellis, P. P. Raychev, and P. A. 
Terziev, Phys. Lett. B 584 (2004) 40. 

\end{thebibliography}
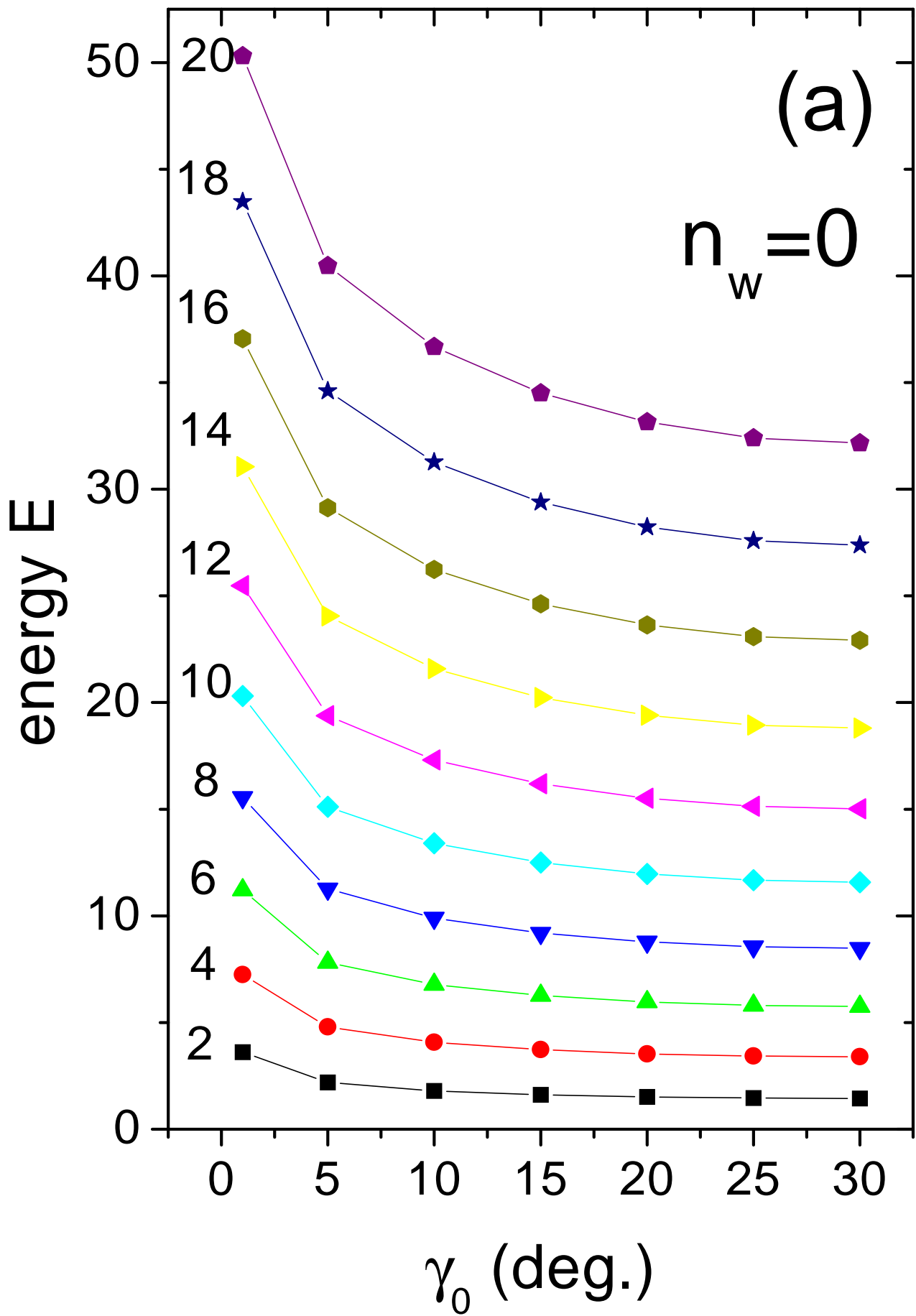
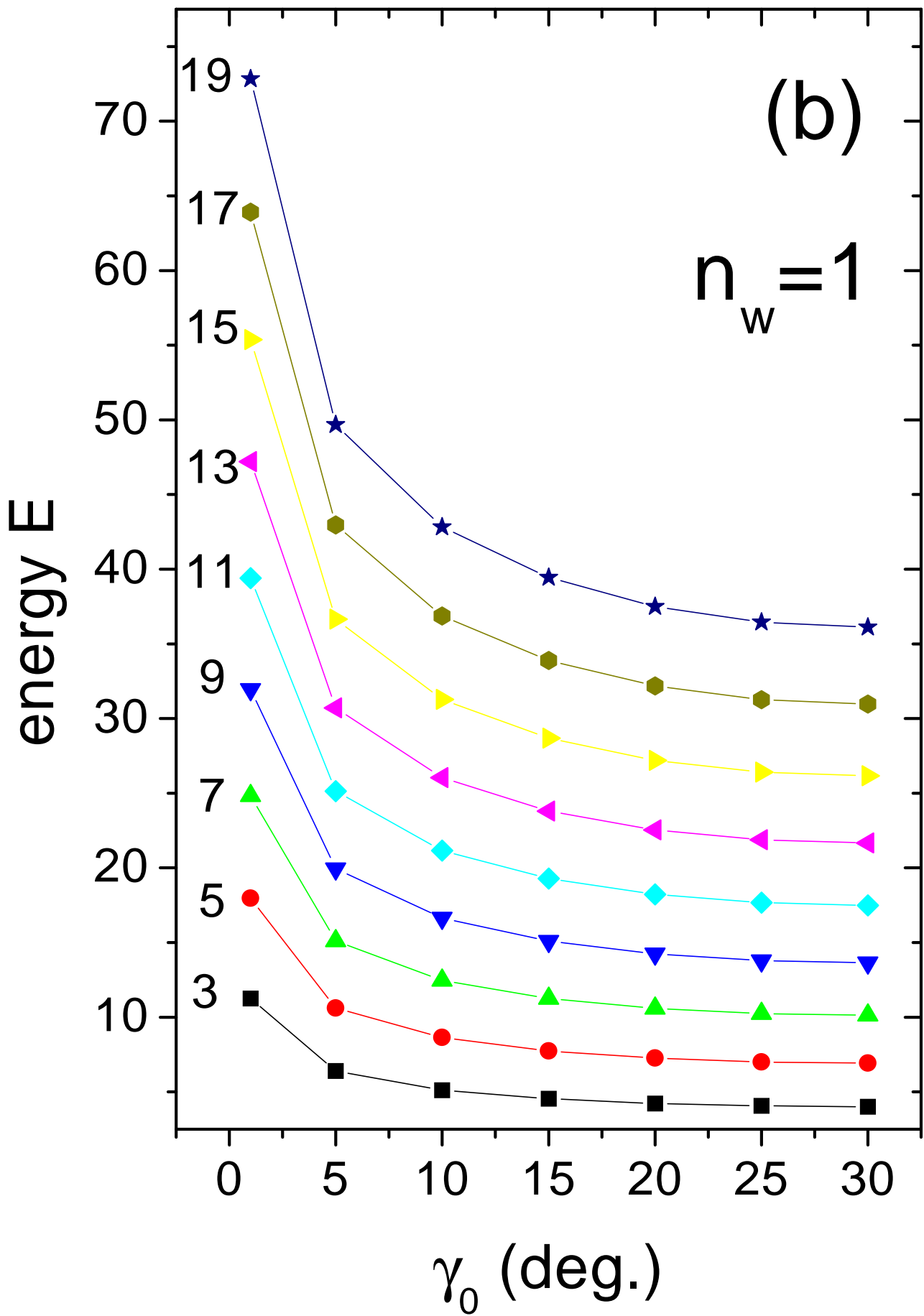
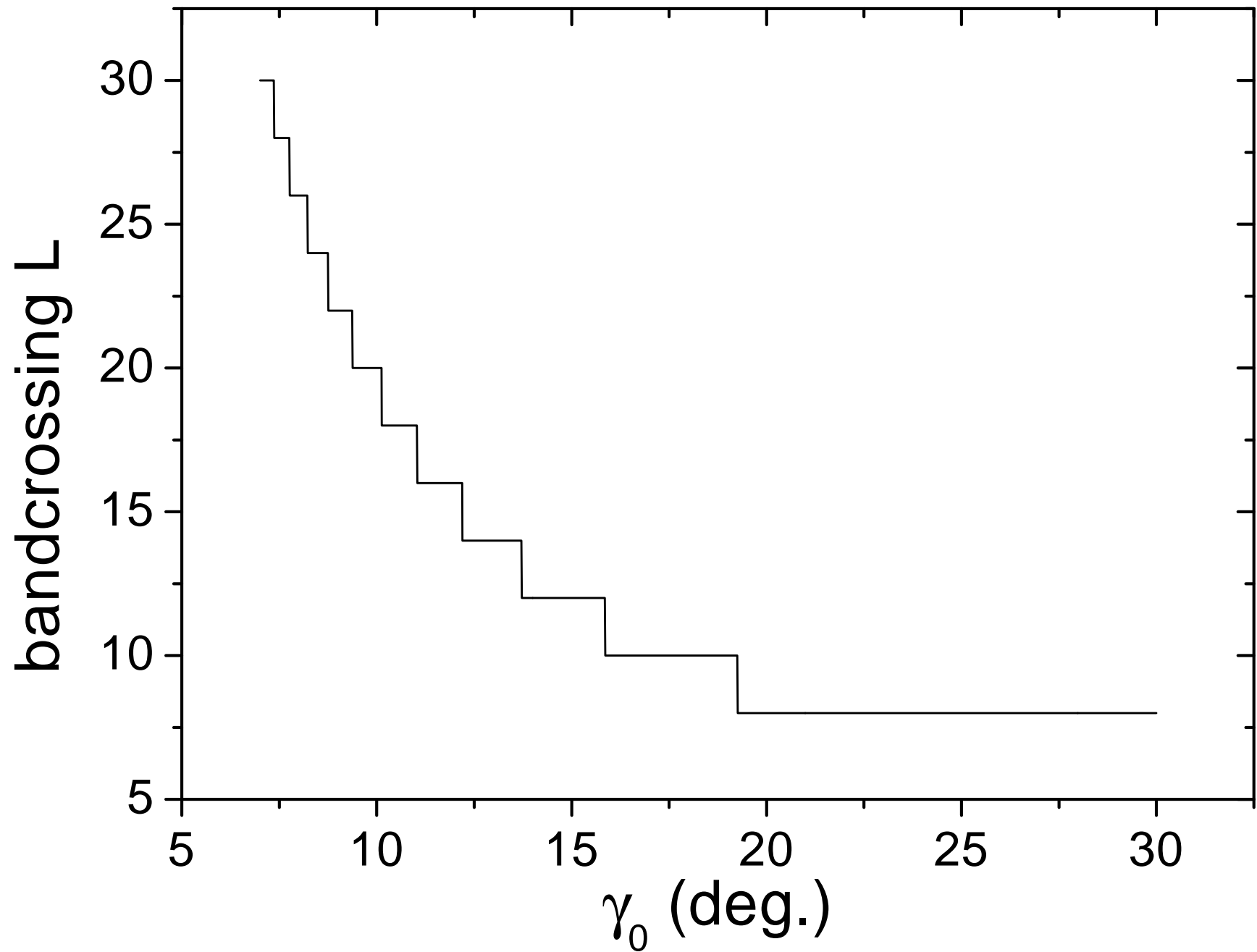
\end{document}